\documentclass[aps,superscriptaddress,showpacs,amsmath,amssymb,amsfonts,altaffillsymbol,twocolumn,notitlepage,prl]{revtex4-1}

\usepackage{graphicx} 
\usepackage{epstopdf}
\usepackage{amsmath}
\usepackage{color}
\usepackage{multirow}
\usepackage{amssymb}
\usepackage{dcolumn}%
\usepackage{bm}%
\usepackage{ctable}
\usepackage{tabularx}
\usepackage{graphicx}

\begin{document}

\begin{titlepage}
\title[]{\Large \bf Yielding behavior of glasses under asymmetric cyclic deformation}

\author{Monoj Adhikari}
\affiliation{Jawaharlal Nehru Centre for Advanced Scientific Research, Jakkar Campus, 560064 Bengaluru, India}
\author{Muhittin Mungan}
\affiliation{Institut f\"{u}r angewandte Mathematik, Universit\"{a}t Bonn, Endenicher Allee 60, 53115 Bonn, Germany}
\author{Srikanth Sastry}
\email[Corresponding author: ]{sastry@jncasr.ac.in}
\affiliation{Jawaharlal Nehru Centre for Advanced Scientific Research, Jakkar Campus, 560064 Bengaluru, India}

\begin{abstract}
We consider the yielding behaviour of a model glass subjected to asymmetric cyclic shear deformation, wherein the applied strain varies between 0 and a maximum value $\gamma_{\rm max}$, and study its dependence on the degree of annealing of the glass and system size. The yielding behaviour of well annealed glasses (unlike poorly annealed glasses) display striking differences from the symmetric case, with the emergence of an intermediate strain regime with substantial plasticity but no yielding. The observed behaviour is satisfactorily captured by a recently proposed model. For larger system sizes, the intermediate strain regime narrows, leading to a remarkable reversal of yield strain with annealing.
\end{abstract}

\maketitle

\end{titlepage}

The response to applied mechanical stresses is a fundamental characteristic of solids that is of  central relevance to their use as structural materials. For large enough applied stresses or deformations, {\it plastic} deformations contribute significantly to such response, leading eventually to yielding and flow. In the case of amorphous solids, ranging from hard glasses such as oxide glasses to soft solids, these plastic deformations are relevant for understanding their yielding behaviour and rheology \cite{schuh2007mechanical,bonn2017yield}. Recent years have witnessed significant activity in developing a statistical mechanical description of these phenomena \cite{falk2011deformation,bonn2017yield,nicolas2018deformation,parmar2020mechanical}. Yielding behaviour in model amorphous solids has been investigated experimentally \cite{nagamanasa2014experimental,keim2013yielding,denisovSR15}, through computer simulations \cite{shi2005strain,shi2007evaluation,fiocco2013oscillatory,regev2013reversibility, wisitsorasak2017dynamical,jaiswal2016mechanical,kawasaki2016macroscopic,regev2015reversibility,jin2018stability,procaccia2017mechanical,leishangthem2017yielding,ozawa2018random,parmar2019strain,barbot2020rejuvenation,bhaumik2021role,yeh2020glass,bhaumiksilica2021,bhaumik2D2021}, and theoretical investigations including the study of elastoplastic models and corresponding mean field theories \cite{parisi2017shear,wisitsorasak2012strength,wisitsorasak2017dynamical,nicolas2018deformation, liu2020oscillatory,ParleyPhyFluids2020,sastry2021models,mungan2021metastability,parley2021}. These investigations have largely focused on the response to uniform shear, but several investigations have explored yielding behaviour under cyclic deformation \cite{fiocco2013oscillatory,regev2013reversibility,priezjev2013,leishangthem2017yielding,parmar2019strain,bhaumik2021role,yeh2020glass,liu2020oscillatory, Maloney2021,bhaumiksilica2021,bhaumik2D2021,sastry2021models,mungan2021metastability,parley2021}.

A specific issue that has received considerable attention recently is the role of annealing of the glasses that are subjected to deformation, in determining the nature of yielding. Indeed, under both uniform shear and cyclic shear, it has been demonstrated that a qualitative change occurs in the yielding behaviour when the degree of annealing of the glasses increases. Under cyclic deformation \cite{bhaumik2021role,yeh2020glass,liu2020oscillatory,bhaumiksilica2021,bhaumik2D2021,sastry2021models,mungan2021metastability,parley2021}, poorly annealed glasses display significant mechanically induced annealing, and converge to a common threshold energy, before yielding takes place. In contrast, well annealed glasses (with energies below the threshold energy), do not display any change in properties with increasing amplitude of shear until the yielding amplitude is reached.  The subsequent yield event is accompanied by a discontinuous change in energy and stress, the amount of which depends on the degree of annealing. Under uniform shear corresponding responses are observed \cite{ozawa2018random,bhaumik2021role}.

Apart from the context of yielding, response to cyclic shear has been 
investigated in order to understand the reversible to irreversible transition, {\em i.e.} the  
transition from a dynamics towards an {\it absorbing} state to one where the dynamics is diffusive, as 
observed in non-Brownian 
colloidal suspensions, glasses, related systems and models thereof  \cite{pine2005chaos,corte2008random,ness2018shaken,nagasawa2019classification,martiniani2019quantifying,das2020unified}. Memory formation in models of suspensions and glasses have also been a subject of considerable interest \cite{adhikari2018memory,fiocco2014encoding,mungan2019networks,keim2019memory}. 

In many of these works, particularly when related to yielding, the cyclic deformations protocols have been {\it symmetric}, i.e. the applied strain of the system is varied through a cycle as $0 \rightarrow \gamma^{\rm sym}_{\rm max} \rightarrow  0 \rightarrow  -\gamma^{\rm sym}_{\rm max} \rightarrow 0$, where $\gamma^{\rm sym}_{\rm max}$ is the amplitude of shear \cite{leishangthem2017yielding,parmar2019strain,bhaumik2021role,yeh2020glass,liu2020oscillatory, Maloney2021}. Given that significant structural change is observed below yielding for poorly annealed cases but not for well annealed cases, one may expect that the choice of range from $\gamma_{\rm min}$ to $\gamma_{\rm max}$ over which the strain is varied cyclically 
may significantly influence the plasticity and yielding behaviour. Indeed, such dependence is of practical importance in determining the characteristics of fatigue and fatigue failure \cite{schuh2007mechanical,suresh1998fatigue}, which in turn 
dictate the scope and limits of operability of such materials in real-life applications.

With the aim of investigating the dependence of the nature of plasticity and yielding on particular cyclic deformation protocols, here we consider the response to 
{\it totally asymmetric} cycles of shear, $0 \rightarrow \gamma^{1}_{\rm max} \rightarrow  0 \rightarrow \gamma^{1}_{\rm max} \dots$. Specifically, we simulate a model glass employing the athermal quasistatic (AQS) shear protocol, and study the response of samples with a widely differing degree of annealing, system size, and subjected to a range of strain  amplitudes $\gamma^{1}_{\rm max}$.

The observed behaviour is found to be markedly different from the case of {\it symmetric} cyclic shear. For poorly annealed glasses (with initial energies above the threshold energy) the yielding behaviour follows the symmetric case, with a rescaling of the strain amplitudes, as we discuss below. 
For the well-annealed samples, for the smaller system sizes considered,  we find an {\it intermediate} range of $\gamma^{1}_{\rm max}$ values over which the  stress  decreases from the maximum value attained, {\it i.e.} {\em beyond} the stress peak, but no diffusive behavior is present. The onset of diffusive behavior, at a larger $\gamma^{1}_{\rm max}$ value, is identified with yielding \cite{fiocco2013oscillatory,leishangthem2017yielding,parmar2019strain,bhaumik2021role}.
In order to better understand our results, we consider a recently proposed mesostate model \cite{sastry2021models} that we adapt to the asymmetric shear protocol. We show that this model qualitatively captures the observed behaviour, thereby shedding light on the underlying relaxation mechanisms. We investigate the dependence on system size $N$, and find that the intermediate window of $\gamma^{1}_{\rm max}$ narrows with system size for well annealed case. Our results suggest, but cannot conclusively demonstrate, that the intermediate strain window will vanish as $N \rightarrow \infty$, but the narrowing itself leads to a remarkable conclusion: Under asymmetric shear, well annealed glasses will yield at smaller strain amplitudes than poorly annealed glasses, reversing the trend observed in the case of symmetric shear protocols.


\noindent{\it Simulations.}
We perform AQS simulations of a three dimensional model glass former, the $80:20$ Kob-Andersen binary Lennard-Jones mixture (KA-BMLJ), in which particles interact with a Lennard-Jones potential,  employing a quadratic cutoff (with details provided in the Supplemental Material (SM)  \footnote{See Supplemental Material at {\tt <URL>} for details on the KA-BMLJ model, strain amplitude and system size dependence for poorly annealed glasses, additional details on the transition to the diffusive state, and the absence of anisotropy of the sheared configurations,  which includes the additional reference \cite{LeesEdwards}}).
The results presented here were obtained from the simulations with system size 
$N =$ $200 (25), 400 (15), 800 (15)$, $4000 (15), 8000 (3), 32000 (2)$, and $ 64000 (1)$ particles, where the numbers in parentheses indicate the number of independent samples. With $V$ being the sample volume, we equilibrate the system at fixed density $\rho = N/V = 1.2$ 
in the liquid state and at reduced temperatures $T=1.0, 0.60, 0.466, 0.40, 0.37$ {\it via} a constant temperature molecular dynamics simulation. The energy of equilibrated configurations are minimized to obtain {\it inherent structure} or {\it glass} configurations, which have average energies per particle $E_{\rm init}=-6.89, -6.92, -6.98,  -7.03, -7.05$, respectively, for the $T$ values indicated. Energy minimization is performed using the conjugate gradient algorithm. The molecular dynamics and AQS simulations are performed using LAMMPS \cite{plimpton1995fast}.

\noindent{\it Shear deformation protocol.}
 The inherent structures, or glasses, are subjected to cyclic shear deformation in the $xz$ plane using the  AQS protocol,  which involves the application of strain by small increments ($d\gamma=2 \times 10^{-4}$) followed by energy minimization (further details in the SM). We apply asymmetric shear cycles that follow the sequence: $0 \rightarrow \gamma^1_{\rm max} \rightarrow  0$, where $\gamma^1_{\rm max}$ is the amplitude of deformation, continuing until a steady state is reached, in which the system either reaches an {\it absorbing} state so that the same sequence of configurations is visited during subsequent cycles, or a  {\it diffusive} state with no periodicity but statistically stationary properties.
 
For symmetric cyclic shear \cite{fiocco2013oscillatory,leishangthem2017yielding}, the {\it yield strain amplitude} $\gamma^{\rm sym}_y$ has been identified as the strain value marking the onset of the {\it diffusive} state, which also coincides with a discontinuous stress drop from the largest stress value attained just prior yielding \cite{fiocco2013oscillatory,leishangthem2017yielding}.
As described below, for asymmetric shear, the location of the stress peak and the onset of diffusion differ in general, and we will therefore distinguish these strain values as $\gamma_{\rm peak}$ and $\gamma_{\rm diff}$, respectively. Note that both for symmetric and asymmetric shear, significant plastic rearrangements occur before an {\it absorbing} steady state is established.

\begin{figure}[htp]
\centering
\includegraphics[width=0.47\textwidth]{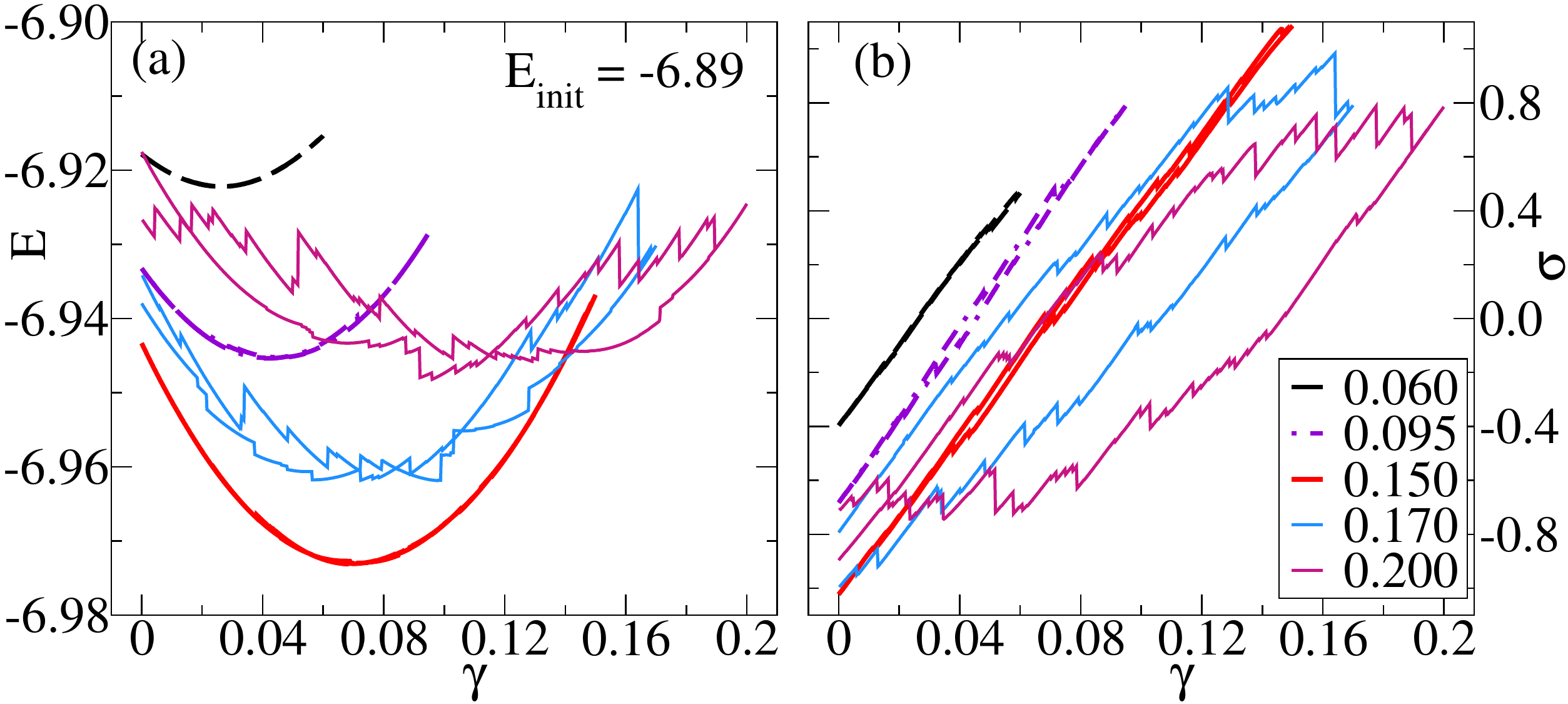}
\caption{Variation of the energy and shear stress through one cycle of strain in the steady state for a single  poorly-annealed glass sample with $N = 4000$ and initial energy $E_{\rm init} = -6.89$. The different curves correspond to different driving amplitudes $\gamma^{1}_{\rm max}$, as indicated in the legend of panel (b).}
\label{fig-stresstrain}
\end{figure}

\noindent{\it Poorly-annealed glasses.} We first consider the steady states attained under asymmetric shear for a poorly annealed glass for $N = 4000$, with $E_{\rm init} = -6.89$.  In Fig. \ref{fig-stresstrain} we consider a single sample and show how the energy (left panel) and stress (right panel), as a function of the applied strain $\gamma$, change over a driving cycle, once a steady-state has been reached (Data averaged over samples is shown in SM Fig. S1). The curves shown correspond to different driving amplitudes $\gamma^1_{\rm max}$, as indicated in (b).
The energy {\it vs.} strain curves display a minimum $E_{\rm min}$ at a non-zero strain value $\gamma_{E_{\rm min}}$ which we denote as the {\it plastic strain}. For $\gamma^{1}_{\rm max} \le 0.15$, $\gamma_{E_{\rm min}}$ increases approximately as $\gamma_{E_{\rm min}} = \gamma^{1}_{\rm max}/2$ (more below).
For $\gamma^{1}_{\rm max} > 0.15$, the energy curves display two well-separated minima and the response ceases to be periodic.
These two observations can be explained by a shift in the plastic strain: 
during the  transient leading to periodic response, the plastic strain evolves towards $\gamma_{E_{\rm min}} = \gamma^{1}_{\rm max}/2$, so that the asymmetric shear deformation protocol  $0 \to \gamma^1_{\rm max} \to 0$, effectively becomes a symmetric one around the plastic strain $\gamma^{1}_{\rm max}/2$ with amplitude $\gamma^1_{\rm max}/2$. In fact, the observation of yielding for $\gamma^{1}_{\rm max} > 0.15$, when taking into account the shift of $\gamma_{E_{\rm min}}$, implies an effective yielding amplitude of about $0.075$ under a symmetric shear protocol, which is consistent with the observed value $\gamma_{y}^{\rm sym} \approx 0.075$ \cite{bhaumik2021role}.  

\begin{figure*}[t!]
\centering 
\includegraphics[width=0.99\textwidth]{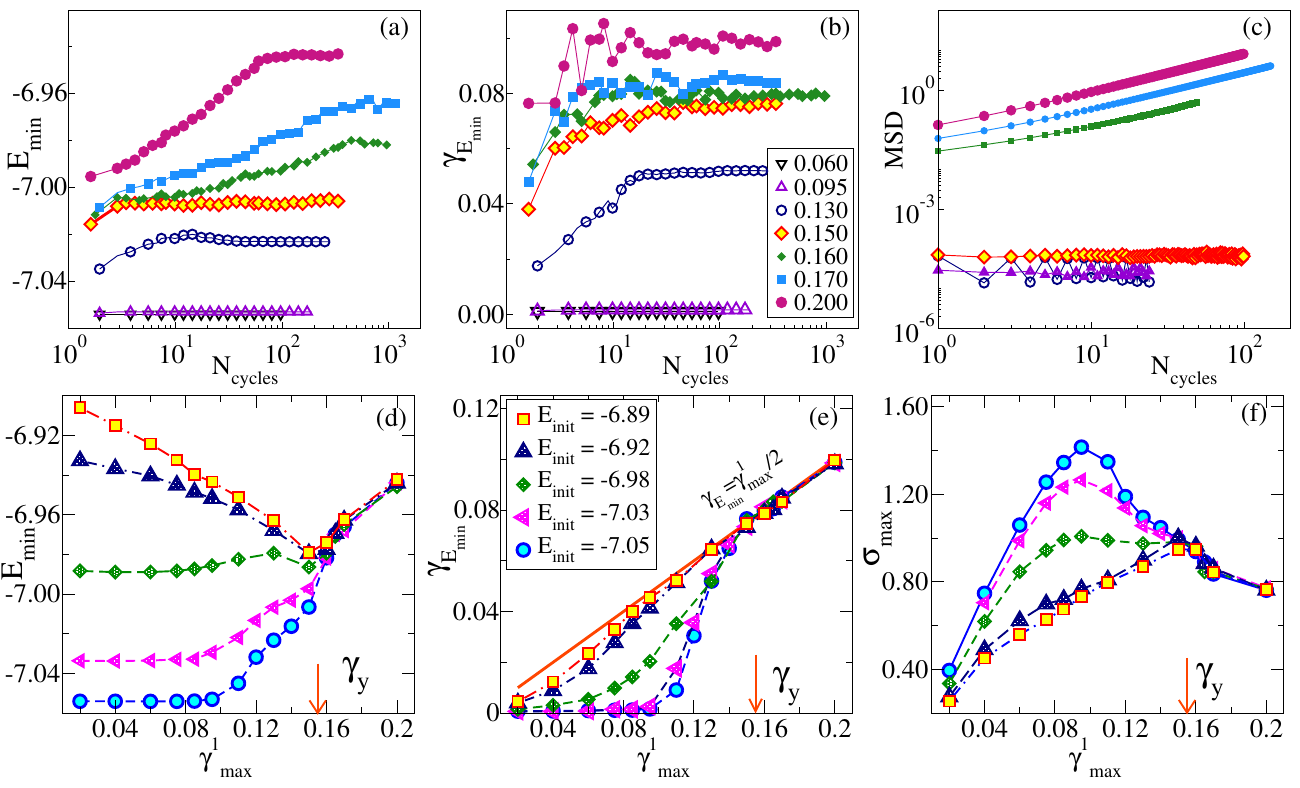}
\caption{Behavior of the minimum energy $E_{\rm min}$ of a cycle (a), the plastic strain $\gamma_{E_{\rm min}}$ at which it is attained (b), and the mean-squared displacement (MSD) from the initial configuration (c),  as a function of cycles of strain $N_{\rm cycles}$. The data shown is for a well-annealed glass with $N = 4000$, initial energy $E_{\rm init} = -7.05$, and driven by a range of asymmetric shear strain amplitudes 
$\gamma^1_{\rm max}$, as indicated in the legend of (b). (d) -- (f): The yielding diagram showing how steady states properties (for $N = 4000$) depend on asymmetric strain amplitude $\gamma^1_{\rm max}$ and the degree of annealing, as indicated in the legend of (e). The steady state values shown are $E_{\rm min}$, $\gamma_{E_{\rm min}}$ and shear stress $\sigma_{\rm max}$ at maximum strain $\gamma^1_{\rm max}$. The yield strain amplitude $\gamma_{y}$, below which the MSD curves in (c) show zero slopes, is marked by red arrows in panels (d) -- (f). 
}
\label{Fig:evo-Annealing}
\end{figure*}

\noindent{\it Well-annealed glasses.} The picture changes dramatically when we consider well-annealed glasses. In the case of symmetric shear, a 
well-annealed glass with $N = 4000$ and $E_{\rm init} = -7.05$ shows, under increasing strain  amplitude $\gamma_{\rm max}^{\rm sym}$, no change in the minimum Energy $E_{min}$ and plastic strain of the cylic response until yielding, which occurs at $\gamma_{\rm max}^{\rm sym} = 0.105$ \cite{bhaumik2021role}. 
Fig. \ref{Fig:evo-Annealing} (a) shows $E_{min}$ as a function of cycle number $N_{\rm cycles}$ for selected asymmetric shear amplitudes $\gamma^{1}_{\rm max}$. As in the case of symmetric shear, for small  $\gamma^{1}_{\rm max} = 0.060, 0.095$ the energy does not change with $N_{\rm cycles}$. However, different from the response under symmetric cyclic shear, for larger values $\gamma^{1}_{\rm max} \ge 0.130$, $E_{min}$ increases with $N_{\rm cycles}$ and saturates at values that grow with $\gamma^{1}_{\rm max}$.
Interestingly, the plastic strain
$\gamma_{E_{min}}$ exhibits a similar behavior, 
Fig.~\ref{Fig:evo-Annealing}(b): $\gamma_{E_{min}}$ remains fixed at $0$ for  $\gamma^{1}_{\rm max} = 0.060, 0.095$, but grows with  $N_{\rm cycles}$ as well as $\gamma^{1}_{\rm max}$ for the larger $\gamma^{1}_{\rm max}$ values. Moreover, the asymptotic behavior $\gamma_{E_{min}} \approx \gamma^{1}_{\rm max}/2$ is attained only when $\gamma^{1}_{\rm max} > 0.150$, while for $\gamma^{1}_{\rm max} = 0.130, 0.150$ the 
asymptotic values of $\gamma^{1}_{\rm max}$ lie between $0$ and $\gamma^{1}_{\rm max}/2$.
Nevertheless, for $\gamma^{1}_{\rm max} > 0.095$, substantial plastic deformations appear to lead to finite  $\gamma_{E_{min}}$ values, even though, as shown in Fig. \ref{Fig:evo-Annealing} (c), a diffusive steady-state is reached only for $\gamma^{1}_{\rm max} > 0.150$. 

The corresponding results for the poorly annealed glasses with $E_{\rm init} = -6.89$ are shown in the SM, displaying a gradual change of both  $E_{min}$ and $\gamma_{E_{min}}$ with  $N_{\rm cycles}$ as well as $\gamma^{1}_{\rm max}$, but with yielding occurring only when $\gamma^{1}_{\rm max} > 0.15$, as in the well annealed case. The SM also contains results to show the absence of anisotropy of the sheared glasses.


\noindent{\it The yielding diagram.} We summarise the results for the full range of annealing of the glasses we considered
in Fig. \ref{Fig:evo-Annealing} (d)-(f). Fig. \ref{Fig:evo-Annealing} (d) shows the steady state energies as a function of $E_{\rm init}$ and the strain amplitude $\gamma^{1}_{\rm max}$. For the poorly-annealed glasses with $E_{\rm init}=-6.89,  -6.92$, the variation of the energies is similar to the case of symmetric shear, displaying a non-monotonic change in energy across the yielding amplitude. In sharp contrast, for the well-annealed glasses, $E_{\rm init} =  -7.03, -7.05$, the energies remain constant up to values of $\gamma^{1}_{\rm max}$ that turn out to be close to the yield amplitudes 
that had been established for the symmetric case as $\gamma_{y}^{\rm sym} = 0.098, 0.105$, respectively \cite{bhaumik2021role}.
For values of $\gamma^{1}_{\rm max}$ beyond this, $E_{\rm min}$ increases with 
$\gamma^{1}_{\rm max}$ until reaching
a value of $E_{\rm min} \approx -6.985$,
which was identified as the {\it threshold energy} in \cite{bhaumik2021role} across which the character of yielding changes in the symmetric case. The intermediate case of $E_{\rm init}=-6.98$ displays an interesting non-monotonic behaviour. 

Irrespective of the degree of annealing, for $\gamma^{1}_{\rm max} > 0.15$ the energies $E_{\rm min}$ all increase along a common curve, and we identify 
$\gamma_{\rm diff} \approx 0.155$ as the onset of the diffusive steady-state regime. A rationalisation of these results can be found in the behaviour of the plastic strain $\gamma_{E_{min}}$ shown in Fig. \ref{Fig:evo-Annealing} (e). The extent to which  $\gamma_{E_{min}} \approx \gamma^{1}_{\rm max}/2$ is achieved can be seen as indicative of the plastic deformations that have taken place before a steady state was reached. Clearly, for  $E_{\rm init} =  -7.03, -7.05$, little deformation occurs until $\gamma^{1}_{\rm max} \approx 0.1$, whereas for higher  $\gamma^{1}_{\rm max}$ the plastic strain shifts, along with the energies, until configurations are reached which are stable under the imposed cyclic strain.  For higher $E_{\rm init}$, such reorganizations occur for all  $\gamma^{1}_{\rm max}$, and to a greater degree for larger $E_{\rm init}$. Finally, in Fig.~\ref{Fig:evo-Annealing}(f) we consider the variation of shear stress $\sigma_{\rm max}$ evaluated at  $\gamma^{1}_{\rm max}$. Once again, for $E_{\rm init}=-6.89, -6.92$, we observe a monotonic increase of $\sigma_{\rm max}$ before yielding, but for lower $E_{\rm init}$, we observe a highly unusual {\it non-monotonic} change of $\sigma_{\rm max}$, well before the yield point. The maximum stress values obtained when $\gamma^{1}_{\rm max} = \gamma_{\rm peak} = 0.095$ are comparable to the yield stress values in the symmetric cyclic shear case \cite{bhaumik2021role}.

Unlike the symmetric shear case, the location of the stress maximum and the onset of diffusive behaviour do not coincide for asymmetric shearing of well annealed glasses,  as seen in Fig.~\ref{Fig:evo-Annealing}(f), (c), with $\gamma_{\rm peak}$ (location of the stress maximum) $< \gamma_{\rm diff}$ (onset of the diffusive steady state). These results are surprising in the context of cyclic shear, since previous observations of plasticity before yielding have invariably been associated with annealing (decrease of energy). The new non-trivial ingredient that is brought forth by the asymmetric shear results is dynamics induced by deformation along the {\it plastic strain} axis. 
We will return to the regime $\gamma_{\rm peak} < \gamma^1_{\rm max} < \gamma_{\rm diff}$ when discussing the effect of system sizes.

\begin{figure}[htp]
\centering
\includegraphics[width=0.49\textwidth]{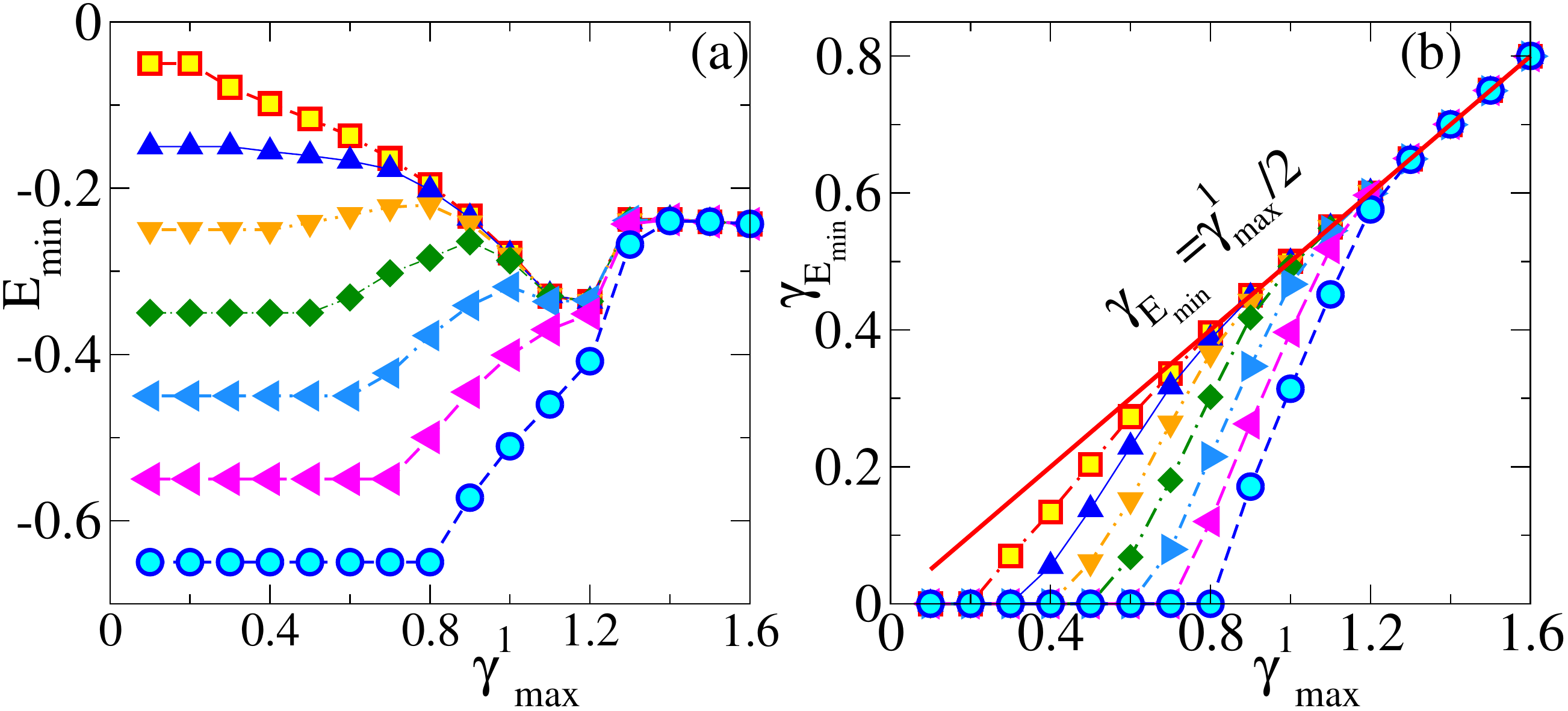}
\caption{The dependence of energy $E_{min}$ and plastic strain $\gamma_{E_{min}}$ as a function of initial energy and strain amplitude for the {\it mesostate} model defined in \cite{sastry2021models}, with a constraint on the magnitude of change in mesostate energy $\delta \epsilon = 0.05$, and refernece strain $\delta \gamma = 0.1$.}
\label{Fig:model}
\end{figure}

\noindent{\it Simulations of a mesostate model.} We next consider whether the observed behaviour can be reproduced by a mesostate model proposed to describe yielding under cyclic deformation, in \cite{sastry2021models}. The model is defined in terms of {\it mesostates}, each of which is characterised by a minimum energy $-\epsilon$, which is  attained at a {\it plastic strain} $\gamma_{\rm min}$, with the total energy of the state at a given strain being given by $E_{\epsilon, \gamma_{\rm min}}(\gamma) = -\epsilon + {\kappa \over 2} (\gamma - \gamma_{\rm min})^2$. The stability range of each mesostate is given by $\gamma^{\pm} = \gamma_{\rm min} \pm \sqrt{\epsilon}$. Upon variation of strain when a given mesostate $(\epsilon, \gamma_{\rm min})$ becomes unstable, a new state $(\epsilon', \gamma'_{\rm min})$ is drawn randomly 
subject to the condition $E_{\epsilon', \gamma'_{\rm min}}(\gamma) < E_{\epsilon, \gamma_{\rm min}} (\gamma)$. We study the model here subject to the additional range restrictions $\epsilon' \in (\epsilon -\delta \epsilon, \epsilon +\delta \epsilon)$, $\gamma'_{\rm min} \in (\gamma_{\rm min} -\delta \gamma, \gamma_{\rm min} +\delta \gamma)$. 
Note that the restrictions imposed on $\epsilon'$ and  $\gamma'_{\rm min}$ (though not the exact values $\delta \epsilon$ and $\delta \gamma$) are found to be important for reproducing the simulation results qualitatively, and we choose $\delta \epsilon = 0.05$, $\delta \gamma = 0.1$.

As shown in  Fig.~\ref{Fig:model}, 
the qualitative aspects of the observations from particle simulations, Fig.~\ref{Fig:evo-Annealing}(d) and (e),  are remarkably well reproduced by the model. 
This suggests, on the one hand, that the mesostate model has the right ingredients to describe yielding in amorphous solids, and on the other hand, that the phenomena we observe are robust and generic. The fact that only some specific choices of model parameters reproduce the simulation results, in particular the evolution of the plastic strain, offers guidance for how such modeling may be pursued to faithfully capture behaviour of amorphous solids subjected to deformation. 

\begin{figure*}[htp]
\centering
\includegraphics[width=0.99\textwidth]{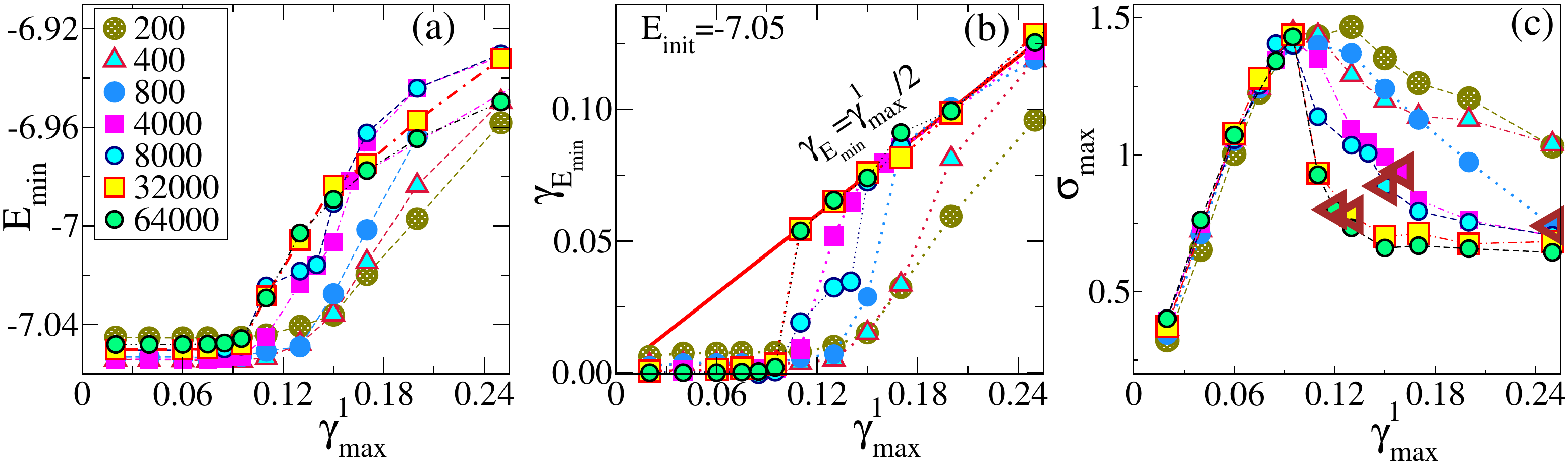}
\caption{System size dependence of $E_{min}$,  $\gamma_{E_{min}}$, $\sigma_{max}$ for the well annealed system. The red line corresponds to $\gamma_{E_{min}}$= $1/2\gamma^{1}_{max}$. The brown left triangles in (c) indicate the smallest $\gamma^1_{\rm max}$ for which the sheared glasses exhibit diffusive behaviour. }
\label{SM:Fig:systemsize}
\end{figure*}

\noindent{\it System size analysis.} In order to interrogate better the intermediate regime $\gamma_{\rm peak} < \gamma^1_{\rm max} < \gamma_{\rm diff}$, that emerges for the well-annealed samples,
we consider next the effect of system size. In  Fig. \ref{SM:Fig:systemsize} (a) and (b) we plot the behavior of the minimum energy $E_{min}$ and the plastic strain $\gamma_{E_{min}}$ at steady state against $\gamma^1_{max}$, for different system sizes, $E_{init}=-7.05$. A strong system size dependence is apparent: the smaller the system size, the larger the strain $\gamma^{1}_{max}$ value beyond which $\gamma_{E_{min}}$= $\gamma^{1}_{max}/2$. In  Fig. \ref{SM:Fig:systemsize} (c) we show the corresponding evolution for $\sigma_{max}$. For system sizes larger than $4000$, the stress maximum appears around a common value $\gamma_{\rm peak} = 0.1$, but the subsequent drop of stress becomes sharper with increasing system size. Although we find $\gamma^1_{max}$ values for each $N$ above $\gamma_{\rm peak}$ but below $\gamma_{\rm diff}$ (see SM for details), we expect that $\gamma_{\rm diff} \rightarrow \gamma_{\rm peak}$ as $N \rightarrow \infty$. This leads to the remarkable conclusion that for asymmetric shear, the yield valuue $\gamma_{\rm diff}$ will be smaller for well annealed glasses compared to poorly annealed glasses, reversing the trend seen for symmetric shear.


In summary, we have investigated the yielding behaviour of a model amorphous solid under asymmetric cyclic shear deformation. We show that such yielding behaviour displays striking new features not observed for symmetric cyclic shear, including the emergence of an intermediate window of strain amplitudes dominated by significant plastic rearrangements and accompanied by a decrease of stress for well annealed glasses. Such a window is expected to vanish for $N \rightarrow \infty$, as our system size results indicate. Nevertheless,  our results reveal the central role played by the non-trivial evolution of 
{\it plastic strain} in the case of asymmetric cyclic shear in determining plastic response, in addition to the evolution of energy. They provide insight into the behaviour of deformed glasses in general.  The qualitative reproduction of these features by a mesostate model for yielding under cyclic shear, point to key aspects that merit attention in building realistic theoretical models of plasticity in amorphous solids on the mesoscale (for which the model is intended and applied here). Investigating finite size effects leads to the remarkable conclusion that for asymmetric shear, the yield strain amplitude is smaller for well annealed glasses, in a reversal of observations for symmetric shear.

 
\paragraph*{Acknowledgements.} 
MM was supported by the Deutsche Forschungsgemeinschaft (DFG, German Research Foundation) under Projektnummer 398962893, the Deutsche Forschungsgemeinschaft (DFG, German Research Foundation) - Projektnummer 211504053 - SFB 1060, and by the Deutsche Forschungsgemeinschaft (DFG, German Research Foundation) under Germany’s Excellence Strategy - GZ 2047/1, Projekt-ID 390685813. We thank the Thematic Unit of Excellence on Computational Materials Science, and the National Supercomputing Mission facility (Param Yukti) at the Jawaharlal Nehru Center for Advanced Scientific Research for computational resources. SS acknowledges support through the JC Bose Fellowship  (JBR/2020/000015) SERB, DST (India).

\bibliography{onesidedmemory}

\end{document}